%% file: NSDI_OAM.tex
\documentclass[pra,superscriptaddress,twocolumn]{revtex4-2}
\usepackage{graphicx, amsmath, braket, color,siunitx,dsfont,lipsum,hyperref}
\usepackage{mathtools,bbold}
\input{SF_Def}

\begin{document}
	\title{Entanglement of Orbital Angular Momentum in Non-Sequential Double Ionization}
	
	\author{Andrew S. Maxwell\orcid{0000-0002-6503-4661}}
	\affiliation{ICFO-Institut de Ciencies Fotoniques, The Barcelona Institute of Science and Technology, 08860 Castelldefels (Barcelona), Spain}
	\affiliation{Department of Physics and Astronomy, Aarhus University, DK-8000 Aarhus C, Denmark}
	
	\author{Lars Bojer Madsen\orcid{0000-0001-7403-2070}}
	\affiliation{Department of Physics and Astronomy, Aarhus University, DK-8000 Aarhus C, Denmark}

	\author{Maciej Lewenstein\orcid{0000-0002-0210-7800}}
	\affiliation{Institut de Ciencies Fotoniques, The Barcelona Institute of Science and Technology, 08860 Castelldefels (Barcelona), Spain}
	\affiliation{ICREA, Pg. Lluís Companys 23, 08010 Barcelona, Spain}

	\begin{abstract}
		We demonstrate entanglement between the orbital angular momentum (OAM) of two photoelectrons ionized via the strongly correlated  process of non-sequential double ionization (NSDI). Due to the quantization of OAM, this entanglement is easily quantified and has a simple physical interpretation in terms of conservation laws. We explore detection by an entanglement witness, decomposable into local measurements, which strongly reduces the difficulty of experimental implementation. We compute the logarithmic negativity measure, which is directly applicable to mixed states, to demonstrate that the entanglement is robust to incoherent effects such as focal averaging. 
		Using the strong-field approximation, we quantify the entanglement for a large range of targets and field parameters, isolating the best targets for experimentalists.
		The methodology presented here provides a general way to use OAM to quantify and, in principle, measure entanglement, that is well-suited to attosecond processes, can enhance our understanding and may be exploited in imaging processes or the generation of OAM-entangled electrons.
	\end{abstract}
	
	\maketitle

	\section{Introduction}
	
	One of the most notorious departures from a classical view of physics is quantum entanglement, a subtle combination of classical correlation \footnote{Classical correlation here refers to two-particle measurement outcomes (e.g. \cite{horodecki_quantum_2009}) and not the interaction between particles as used in electronic structure computations.} and quantum superposition, which not only caused a seismic shift in how the world is viewed, but also provides a resource for quantum computation and metrology \cite{horodecki_quantum_2009,giovannetti_advances_2011,degen_quantum_2017,bongs_taking_2019}. This includes the possibility of entanglement enhanced imaging processes \cite{abouraddy_role_2001,steinlechner_quantumdense_2013,wolfgramm_entanglementenhanced_2013, defienne_polarization_2021}. However, the potential for entanglement to optimize or improve attosecond ($10^{-18}$s) imaging processes is unexplored, and therefore the role and physical insight afforded by entanglement for such processes remains unclear.
	
	Attosecond and strong-field physics deal with processes in matter on the scale of attoseconds \cite{corkum_attosecond_2007,krausz_attosecond_2009}. The promise of resolving atomic and molecular electron dynamics on its natural timescale has led to the development of a host of imaging procedures boasting attosecond time resolution; including strong-field initiated methods like high-order high harmonic spectroscopy \cite{itatani_tomographic_2004,marangos_development_2016}, laser-induced electron diffraction \cite{zuo_laserinduced_1996,niikura_sublasercycle_2002}, photoelectron holography \cite{huismans_timeresolved_2010,figueirademorissonfaria_it_2020} and attosecond pump-probe techniques like attosecond streaking \cite{itatani_attosecond_2002,hentschel_attosecond_2001} and reconstruction of attosecond harmonic beating by interference of two-photon transitions \cite{paul_observation_2001,muller_reconstruction_2002}.
	Although most of the above-stated imaging protocols depend on quantum processes like tunneling or interference, none of them explicitly exploit entanglement.
	
    Following a trend of non-classicality, the inherent quantum nature of attosecond and strong-field physics was revealed recently through the generation of highly non-classical states of light in the driving field for high-harmonic generation and above-threshold ionization \cite{lewenstein_generation_2021, rivera-dean_quantum_2021}.
    Similarly, there has been a steady and growing interest in the role of entanglement in attosecond processes. As early as 1994, a correlation measure, which provides an indication of entanglement based on Schmidt decomposition, was developed \cite{grobe_measure_1994}. 
    Many studies have focused on entanglement between photoelectrons and ions \cite{spanner_entanglement_2007, spanner_coherent_2007,czirjak_emergence_2013,majorosi_quantum_2017,ruberti_quantum_2021,vrakking_control_2021,koll_experimental_2021}, which is an essential part of understanding decoherence \cite{rohringer_multichannel_2009},
    while other studies have focused on electron--electron entanglement \cite{liu_correlation_1999,christov_phaselocking_2019,omiste_effects_2019}. However, all these previous studies involved the calculation of a continuous variable density matrix and the consideration of entanglement measures such as the purity \cite{czirjak_emergence_2013,christov_phaselocking_2019,omiste_effects_2019,ruberti_quantum_2021,vrakking_control_2021}, von Neumann entropy \cite{czirjak_emergence_2013,majorosi_quantum_2017,omiste_effects_2019,ruberti_quantum_2021} or a Schmidt decomposition based measure \cite{grobe_measure_1994,liu_correlation_1999}. These quantities, when derived from continuous variables, have some drawbacks: (i) They are a considerable challenge to compute, and often approximations or restrictions must be imposed, such as one-dimensional calculations. Furthermore, these methods are restricted to pure states (without non-trivial extensions \cite{horodecki_quantum_2009}), while in strong-field experiments we must always consider incoherent effects, for example associated with focal volume averaging, leading to mixed states. (ii) The physical interpretation of these quantities can be difficult, and these measures may not improve understanding of the process or lead to any further applications. (iii) Direct experimental evidence of the entanglement is often practically impossible. In the case of continuous variable entanglement between photoelectrons, it requires the measurement of incompatible observables such as momentum and position.
    
    One solution to these difficulties is to use quantized observables, drastically reducing the required dimensionality. All free particles have such a quantum observable in the form of orbital angular momentum (OAM) \cite{allen_orbital_1992,bliokh_semiclassical_2007}, which manifests by a rotating wavefront of phase. Entanglement of OAM in electron vortex beams has been theoretically discussed in Ref.~\cite{ivanov_creation_2012}. Photons and electrons carrying OAM have recently received a great deal of attention in attosecond physics, for example using infrared (IR) laser fields carrying OAM to produce extreme ultraviolet (XUV) high-order harmonics with OAM, see e.g., Refs.~\cite{zurch_strongfield_2012,hernandez-garcia_attosecond_2013,gariepy_creating_2014,hernandez-garcia_quantumpath_2015,geneaux_synthesis_2016}. 
    Strong-field studies on OAM in photoelectrons include theoretically achieving high OAM values for quasi-relativistic field intensities \cite{velez_generation_2018a} and terahertz fields \cite{cajiaovelez_generation_2020}, exploiting OAM in rescattering electrons to probe bound state structures \cite{tolstikhin_strongfield_2019} and recent work \cite{maxwell_manipulating_2021}, providing new insight to interference vortices \cite{ngokodjiokap_electron_2015} as an interference between pairs of OAM components and \cite{kang_conservation_2021}, where conservation laws for OAM in strong-field ionization were derived. Of particular relevance, is the conservation between the initial quantum magnetic number and final OAM, which occurs for systems with rotational symmetry around the quantization axis \footnote{For linearly polarized light this corresponds to the polarization axis defining the OAM axis.}. Such conservation may be exploited in strongly-correlated two-electron processes, where entanglement could allow for enhanced photoelectron imaging. 

    \begin{figure*}
		\centering
		\includegraphics[width=1.0\textwidth]{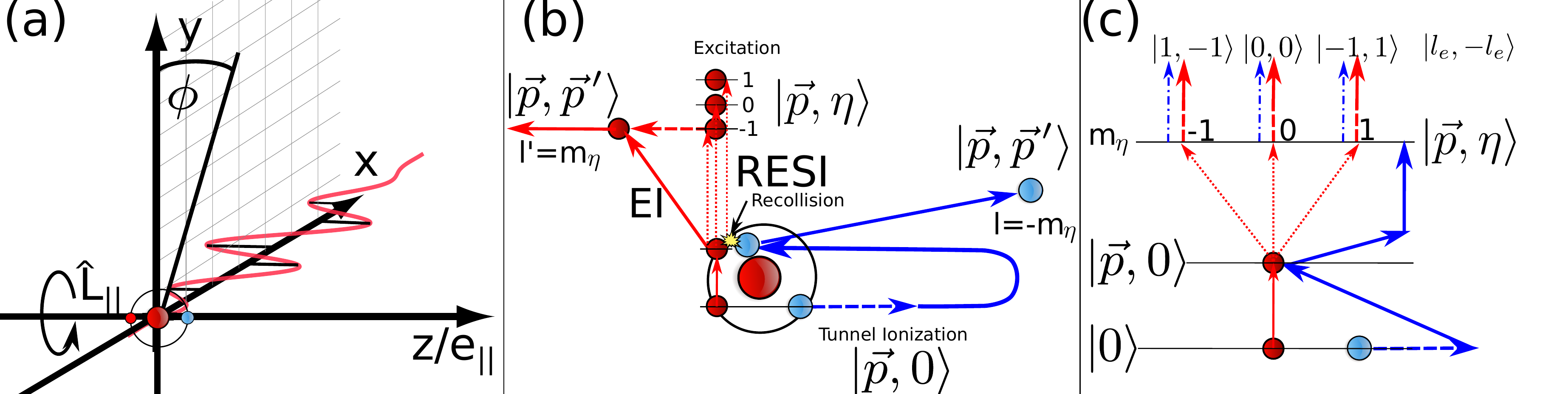}
		\caption{(a) Orientation of the linearly polarized laser, with polarization $e_{||}$ along the $z$ axis taken for the OAM denoted by $\hat{L}_z$ and the target atom. 
		(b) NSDI process depicted for the EI and RESI mechanisms. Interaction via the field (e.g. tunnel ionization) is depicted by a dashed line, excitation in the singly charged ion is denoted by dotted lines, while the recollision and OAM sharing is denoted by the yellow `spark'. The two electron states are defined in the text, following the convention of Ref.~\cite{amini_symphony_2019}. 
		(c) The excitation pathways in RESI, which lead to different final OAM states and an entangled superposition. The final states are given by OAM states, $\ket{l_e,-l_e}$ as used in \eqref{eq:Density_OAM}. The dashed-dotted lines are to denote, which two-electron state the `first' electron ends up in.
		The numbers $-1$, $0$, $1$ refer to the values of the quantum magnetic number $m_{\eta}$ of the intermediate excited state, $\ket{\pv,\eta}$.
		}
		\label{fig:NSDI_process}
	\end{figure*}
	
	Non-sequential double ionization (NSDI) is a highly correlated two-electron ionization process. The details of which  are depicted in \figref{fig:NSDI_process}. The process follows the three-step mechanism [panel (b)] \cite{corkum_plasma_1993}: (i) The first electron is removed from the two-electron ground state $\ket{0}$ by the laser via tunnel ionization into the state $\ket{\pv,0}$ (one electron in the continuum and the other in its ground state). (ii) The continuum electron subsequently undergoes a laser driven recollision with its parent ion. (iii) The energy imparted by the collision allows for two pathways, a second electron is directly ionized in the electron impact (EI) mechanism, or the second electron is excited, resulting in the state $\ket{\pv,\eta}$---here $\eta$ is used to label the excited state---and the second electron subsequently ionizes due to the laser field in the recollision with subsequent ionization (RESI) mechanism. In both cases, the final state of the photoelectrons is $\ket{\pv,\pvp}$, the two electron continuum state. For more information on this notation, see Ref.~\cite{amini_symphony_2019}, while for reviews of NSDI and these mechanisms, see Refs.~\cite{figueirademorissonfaria_electronelectron_2011,becker_theories_2012}.

	Despite strong electron--electron correlation and rescattering being confirmed in NSDI as early as 2000 \cite{weber_correlated_2000,moshammer_momentum_2000}, there has been little focus on the quantum entanglement between the two electrons. 	
	This is primarily for the following reasons, (i) classical models have been very successful in modelling NSDI \cite{panfili_comparing_2001}, (ii) early work suggested entanglement would not play a decisive role.
	A study of correlation in NSDI \cite{liu_correlation_1999}, demonstrated a small degree of `quantum correlation' via a one-dimensional solution of the time-dependent Schrödinger equation and computing a measure related to entanglement, via Schmidt analysis \cite{grobe_measure_1994}, while the findings reported in Ref.~\cite{ho_nonsequential_2005} suggested that classical correlation was sufficient for field intensities greater than $10^{14}$~W/cm$^2$.
	(iii) Furthermore, computation of NSDI is a very arduous task and computation of continuous variable entanglement measures, for the reasons stated above, is even more difficult, while as previously stated, it is unclear if knowledge of the entanglement could add any insight, be useful or measurable in experiment.
	
	In this work we address (ii) and (iii), by exploiting the quantized nature of OAM to clearly demonstrate the manifestation of entanglement in NSDI, which we show may occur only through the RESI pathway via the superposition of excited states.
	The use of a quantized degree of freedom enables a very simple analysis through the logarithmic negativity \cite{vidal_computable_2002} and entanglement witnesses \cite{guhne_experimental_2003,bourennane_experimental_2004,chruscinski_entanglement_2014}, which enables the inclusion of incoherent effects via mixed states from the very beginning, as well as a search over a wide range of the parameters to discover the field and target parameters that lead to the most entangled photoelectron pairs. 
	The interplay of channels of excitation allows photoelectrons to approach maximally entangled states for some final momenta, which could be investigated as a source of OAM entangled electrons.
	We show that the entanglement exhibited is very robust to incoherent averaging of laser intensities over the focal volume. In fact, we find the entanglement does not significantly change, while interference features in NSDI \cite{hao_quantum_2014,maxwell_quantum_2015,maxwell_controlling_2016,quan_quantum_2017}, linked to superposition of channels of excitation, become blurred.
	We demonstrate that this entanglement could be measured by decomposing the entanglement witness. This strongly reduces the difficulty of detecting entanglement by avoiding full tomographic measurements. Furthermore, the OAM entanglement could be used to perform correlated OAM and momentum measurement on the two electrons, providing more information on the target. Thus, paving the way for a new kind of entanglement enhanced attosecond  imaging technique.
	
	The paper is organized as follows. In \secref{sec:Results:NSDI}, we introduce orbital angular momentum (OAM) in non-sequential double ionization (NSDI) and how this process leads to entanglement between the photoelectrons when measured in the quantized OAM basis. Next, in \secref{sec:Results:Measure} we introduce the logarithmic negativity measure and an entanglement witness, and in \secref{sec:Results:Search} we go on to use these measures to quantify the entanglement for a range of field parameters and targets. In \secref{sec:Results:Experimental}, we consider incoherent effect via focal averaging and decomposition of the entanglement witness into local measurements. Finally, in \secref{sec:Discussion}, we state our conclusions and discuss the implementation of OAM measurements and the entanglement witness for NSDI in experiment.
	

	\section{Orbital angular momentum in non-sequential double ionization}
	\label{sec:Results:NSDI}

	A pictorial depiction of non-sequential double ionization (NSDI) is given is \figref{fig:NSDI_process}. Both the direct, electron impact (EI) and recollision excitation with subsequent ionization are shown. There is energy and momentum sharing between the electron during the recollision. In RESI, the extra tunneling step is denoted by dashed lines. The laser field is polarized in the $z$ direction, in the same direction as the total OAM operator $\hat{L}_{||}$, so  the laser field can not change the total OAM since $[\hat{L}_{||}, \hat{H}(t)]=0$, where $\hat{H}(t)$ is the total Hamiltonian of the system, see Refs.~\cite{maxwell_manipulating_2021,kang_conservation_2021} or \appref{sec:Methods:Models:Conservation} for more details.
    
    We will expand the NSDI two-electron continuum wave function in a basis of electron vortex states, the one-electron vortex state is denoted $\ket{\pb,l_e}$ and a plane wave by $\ket{\pv}$. Note, we reserve bold face momentum for the two-dimensional vectors $\pb=(p_{||},p_{\perp})$ and arrow notation, is reserved for three-dimensional vectors associated with a plane wave $\pv=(p_{||},p_{\perp},\phi)$, written in cylindrical coordinates.
    Here, $p_{\perp}=\sqrt{p_x^2+p_y^2}$ is the radial coordinate, $p_{||}=p_{z}$ is the momentum coordinate along the cylindrical axis (parallel to the laser field polarization) and $\phi$ the azimuthal angle $\phi=\arctan(p_y/p_x)$, while $l_e$ is the topological charge or azimuthal OAM. 
    We will employ atomic units throughout, unless otherwise stated.
    The spatial representation of the vortex state is given by \cite{lloyd_electron_2017}
    \begin{align}
    	\braket{\rv|\pb,l_e}&=\frac{1}{(2\pi)^{3/2}}J_{l_e}(p_{\perp} r_{\perp}) e^{i\phi l_e} e^{i p_{||} r_{||}},
    	\intertext{while the momentum representation is}
    	\braket{\kv|\pb,l_e}&=\frac{i^{-l_e}e^{i \phi'l_e}}{2\pi p_{\perp}}\delta(k_{\parallel}-p_{\parallel}) \delta(k_{\perp}-p_{\perp}).
    \end{align}
    We assume in the asymptotic limit of large distances from the atomic nucleus that the two electrons can be written in a basis of a product of vortex states
    \begin{equation}
    	\ket{\pb, l_e, \pb', l_e'}=\ket{\pb, l_e}\otimes \ket{\pb', l^{\prime}_e}.
    \end{equation}
    Typically for NSDI we consider momentum measurement and the associated transition amplitude
    \begin{align}
    M(\pv,\pvp)=\lim\limits_{t\rightarrow \infty} \braket{\pv,\pvp|\psi(t)},
    \end{align}
    where $\ket{\psi(t)}$ denotes the continuum two-electron wavepacket after the interaction with the external field, and $\ket{\pv,\pvp}$ denotes the scattering state with asymptotic momenta $\pv$ and $\pvp$.
    Here, however, we will consider the transition amplitude corresponding to OAM measurement, including only the doubly ionized portion of the system, which may be expressed as
	\begin{align}
		M_{l_e,l_e'}(\pb,\pb')=\lim\limits_{t\rightarrow \infty} \braket{\pb,l_e,\pb',l_e'|\psi(t)},
	\end{align}
    where $M_{l_e,l_e'}(\pb,\pb')$ is the two-dimensional Fourier series coefficients of $M(\pv,\pvp)$
    \begin{align}
    	M_{l_e,l_e'}(\pb,\pb')= \frac{i^{l_e+l_e'}}{(2\pi)^2}\iint \dd \phi \dd \phi' e^{-i\phi l_e-i\phi'l_e'}M(\pv,\pvp).
    	\label{eq:2dFourier}
    \end{align}    
     Note, we will always construct $M_{l_e,l_e'}(\pb,\pb')$ such that electron indistinguishability is accounted for, see \appref{sec:Methods:Models:SFA}.

	The total azimuthal OAM of the two electrons will be conserved at all times
	\begin{equation}
	    l_e+l'_e=m+m',
	\end{equation}
	where $l_e$ and $l'_e$ are the azimuthal OAM of the final vortex state and $m$ and $m'$ are the initial quantum magnetic numbers of the two-electron ground state $\ket{0}$. For the recolliding electron in NSDI $m\ne0$ is strongly suppressed, so we consider $m=0$. Similarly, for the second electron contributions from $m'$ with opposite signs will destructively interfere in the cases we consider, thus we take $m'=0$. 
	The OAM conservation is now trivially $l'_e=-l_e$, see \figref{fig:NSDI_process}(b) and (c).
	For the RESI mechanism, the second electron leaves from an excited state $\ket{\eta}$, and thus we have the additional conservation $l'_e=m_{\eta}$, thus $l_e=-m_{\eta}$, which again can be seen in \figref{fig:NSDI_process}(b) and (c). In \appref{sec:Methods:Models:Conservation}, we derive all these conservation laws explicitly.
	
	Due to the recollision there is OAM sharing and electron correlation in NSDI, while, in the RESI mechanism, the OAM is tunable via the excited state. Furthermore, the excited electron may occupy a superposition of states \cite{hao_quantum_2014,maxwell_quantum_2015,maxwell_controlling_2016,quan_quantum_2017}, which means entanglement can emerge in the OAM degree of freedom. From this point onwards, we will only consider the excited state populates  $m_{\eta}=0,\pm 1$. Values outside this range may occur but can safely be neglected by ignoring photoelectrons with OAM $|l_e|>1$. This reduction on the OAM space will still capture all necessary physics, while limiting the complexity of measurement/ implementation.
	The final  two-electron continuum state corresponding to this scenario can be described by,
	\begin{align}
    	&\ket{\psi}=
    		\iint \dd^2 \pb \dd^2 \pb' \left(
    			M_{1-1}(\pb,\pb')\ket{\pb,1,\pb',-1}\right.\notag\\
    			&+\left.M_{00}(\pb,\pb')\ket{\pb,0,\pb',0}
    			+M_{-11}(\pb,\pb')\ket{\pb,-1,\pb',1}\right),
    	\label{eq:OAM_Qutrit}
    \end{align}
    where we consider a time after the end of the pulse and have suppressed $t$ in our notation for convenience, and we will follow this convention throughout the remainder of this work.
    Here, $\ket{\psi}$ is a maximally entangled qutrit if $M_{1-1}(\pb,\pb')
    =M_{00}(\pb,\pb')
    =M_{-11}(\pb,\pb')
    $. In fact, even for arbitrary $M$'s, it is possible to show, see \appref{sec:Methods:Entanglement:PPT}, that after the momentum coordinates are traced out the resulting mixed state is always entangled given that $\iint \dd^2 \pb \dd^2 \pb' M_{k-k}(\pb,\pb') M^{*}_{k'-k'}(\pb,\pb')\ne0$ $\forall$ $k, k' \in [-1,1]$. This is a strong statement, as it means OAM entanglement can survive integration over all the momentum coordinates, as long as the probability of excitation to states with two or more values of $m_{\eta}$ is non-zero.
    
    \section{Entanglement measure and witness}
    \label{sec:Results:Measure}

	In order to quantify and measure entanglement, we consider the density matrix, $\rho=\ket{\psi}\bra{\psi}$.
	To greatly simplify matters we will assume that the experimentalists are only interested in measuring the OAM, thus we will compute the reduced density matrix, tracing over all the continuous momentum components, see  \eqref{eq:Density_Trace}, leaving an entangled mixed state,
	\begin{align}
	\rho_{OAM}&=\sum_{l_e,l_e'}
	\alpha_{l_e l_e'}
	\ket{l_e,-l_e}\bra{l_e',-l_e'},
	\label{eq:Density_OAM}
	\intertext{with}
	\alpha_{l_e l_e'}&=\int\dd^2 \pb \int \dd^2 \pb'M_{l_e-l_e}(\pb,\pb')M_{{l_e}' -l_e'}^*(\pb,\pb')\notag.
	\end{align}
    Note, for all computations the density matrix will be normalized by its trace. For a derivation, see \appref{sec:Methods:Entanglement:Density}.
	
	\begin{figure*}
		\centering
		\includegraphics[width=0.975\textwidth]{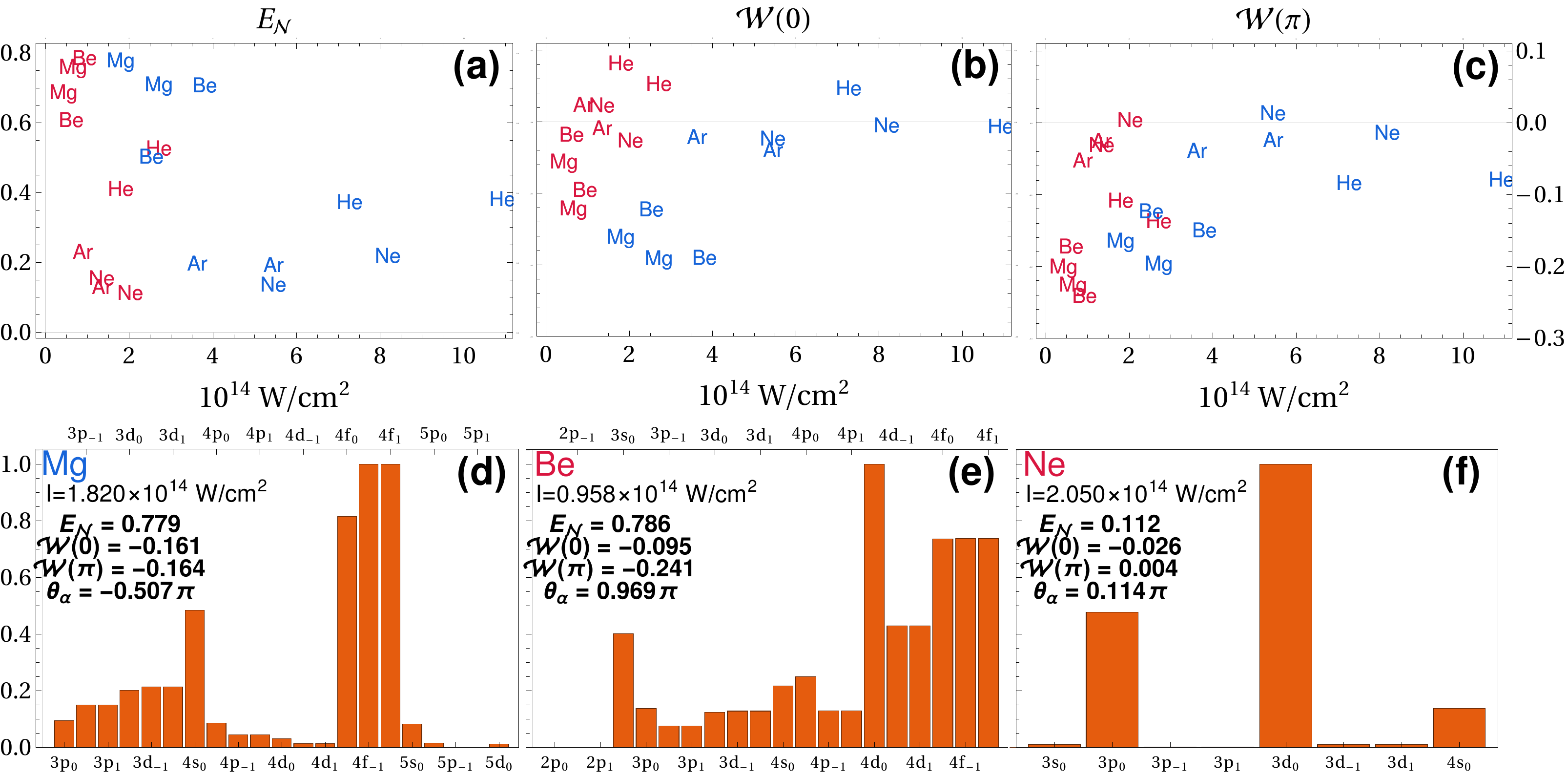}
		\caption{We show an extensive search over targets and parameters. We have quantified the entanglement using the logarithmic negativity ($E_\mathcal{N}$ of \eqref{eq:LogNeg}). The expectation value of the entanglement witnesses ($\mathcal{W}(0)$ and $\mathcal{W}(\pi)$ see \eqref{eq:Witness}) is shown.  Blue text indicates $\lambda=\SI{400}{\nm}$ and red $\lambda=\SI{800}{\nm}$. In the bottom row, for three extremal cases (Mg, Be and Ne at the field values given on the panels), the contribution of intermediate excited states is estimated. This is computed by taking the peak value of the momentum dependent probability of each channel individually. 
		On the bottom panels we denote $E_{\mathcal{N}}$, $\mathcal{W}(0)$, $\mathcal{W}(\pi)$ and the phase $\theta_{\alpha}=\arg(\alpha_{10})$, see \eqref{eq:Density_OAM}. 
		The excited state labels, at the top and bottom of the panels, are in the format $\ket{\eta}=\ket{n\ell_{\scalebox{0.95}{$m_\eta$}}}$, where $n$ is the principle quantum number, $\ell$ is the angular quantum number and $m_\eta$ is the magnetic quantum number.
		}
		\label{fig:EntanglementSearch}
	\end{figure*}

	
	The logarithmic negativity \cite{vidal_computable_2002} is a measure of entanglement, that is valid for mixed state systems, which exploits the positive partial transpose (PPT) \cite{peres_separability_1996,horodecki_separability_1997}. This is ideal for our system as in all cases we can show the electrons are always PPT entangled (see \appref{sec:Methods:Entanglement:PPT}) and it is easy to compute. The logarithmic negativity is given by
	\begin{align}
		E_{\mathcal{N}}=\log_2\left[|| \rho^{T_A}_{OAM} ||_1\right],
		\label{eq:LogNeg}
		\intertext{where the trace norm reads}
		||\rho||_1 =\mathrm{Tr}\left[ \sqrt{\rho^{\dagger}\rho} \right] \notag
	\end{align}
	and $\rho^{T_A}$ is the partial transpose, i.e., the transpose with respect to one subsystem (one of the electrons).
	The logarithmic negativity may vary between 0 and $\log_2(3)\approx1.58$ for the entangled qutrit system considered in \eqref{eq:OAM_Qutrit}. 
	We explore the logarithmic negativity and this upper bound in more detail in \appref{sec:Methods:Entanglement:LogNeg}. The logarithmic negativity is used to quantify entanglement in order to find targets and field parameters, which maximize it. It is not something that can be easily measured but can be used as a theoretical analysis tool providing new insight.
	
	
	To measure if there is entanglement, one approach is to use an entanglement witness, which can distinguish a subset of entangled states as non-separable. They can commonly be associated with an observable for which the expectation value is negative for some entangled states, i.e., for a witness $\mathcal{W}$ and state $\rho$, the condition
	$\mathrm{Tr}\left[\mathcal{W} \rho \right]<0$ implies $\rho$ is entangled.	
	It can be shown that 
	\begin{align}
	\mathcal{W}(\theta)=\frac{1}{d}\mathds{1}-\ket{\nu(\theta)}\bra{\nu(\theta)}
	\label{eq:Witness}
	\end{align}
	with
	\begin{align}
	\ket{\nu(\theta)}=\frac{1}{\sqrt{d}}\sum_{l_e=-1}^1 e^{i \theta l_e}\ket{l_e,-l_e},
	\end{align}
	is a valid entanglement witness, see \appref{sec:Methods:Entanglement:Witness}. Here, $d$ is the dimension of the system, which, as given in \eqref{eq:OAM_Qutrit}, is \num{3}.
	This witness is useful as the state $\ket{\nu(\theta)}$ mirrors the anti-correlated ($l_e'=-l_e$) entangled superposition of the OAM and also contains a system-dependent tuning parameter $\theta$, which can be set depending on the target and field parameters to enhance the detectability. 
	The parameter $\theta$ can be tuned depending on the phase information in the state, in particular depending on $\theta_{\alpha}:=\arg(\alpha_{01})$,
	which represents the  phase between $\ket{0,0}$ and $\ket{\pm1,\mp1}$ in the density matrix $\rho_{OAM}$ given by \eqref{eq:Density_OAM}.
	We will set $\theta$ to two extremes $\theta=0$ or $\pi$ and discuss the correspondence with $\theta_{\alpha}$.
    
    \section{Optimizing Entanglement}
    \label{sec:Results:Search}
    
	Utilizing the logarithmic negativity and entanglement witness, we compute these quantities for many targets and laser parameters and show the results in \figref{fig:EntanglementSearch}. We do this using the strong-field approximation (SFA), the details of which are described in \appref{sec:Methods:Models:SFA} and Refs.~\cite{maxwell_quantum_2015,maxwell_controlling_2016}. The SFA is a simple and approximate method, but it captures the basic OAM correlation and is rapid to compute, so is well-suited to a broad parameter search.
	The parameters are chosen such that the return energy of the first electron $\approx3.17 \up$ is not above the ionization potential of the second, to ensure that the RESI mechanism is dominant. Here, $\up$ is the ponderomotive or quiver energy of the electron in the laser field. Also, we stay approximately in the tunneling regime, with the Kelydsh \cite{keldysh_ionization_1965} parameter $\gamma=\sqrt{\ip/(2\up)}$ spanning 0.87--1.20. 

	Immediately from \figref{fig:EntanglementSearch} (a), it is clear that magnesium and beryllium have the highest logarithmic negativity, while the nobel gases argon and neon have the lowest. In the bottom row, panels (d)--(f), the contribution of channels via different excited states is estimated for the two highest and lowest cases. In the case of the nobel gases, states with $m_{\eta}=0$ dominate, thus reducing the entanglement as a single OAM prevails, see panel (f). We have verified that this selectivity occurs during the recollision: For nobel gases, the second electron initial $p$-state (with $m'=0$) is aligned along the OAM axis and transitions to excited states which keep this alignment are more probable. 	
	For magnesium and beryllium in panels (d) and (e), there is a more balanced superposition across $m_{\eta}=-1,0,1$, the initial $s$-state of these targets is spherically symmetric, so there will be no directional preference for the excited state, leading to higher logarithmic negativity and a higher degree of entanglement.
	
	\begin{figure*}
		\centering
		\includegraphics[width=1.01\textwidth]{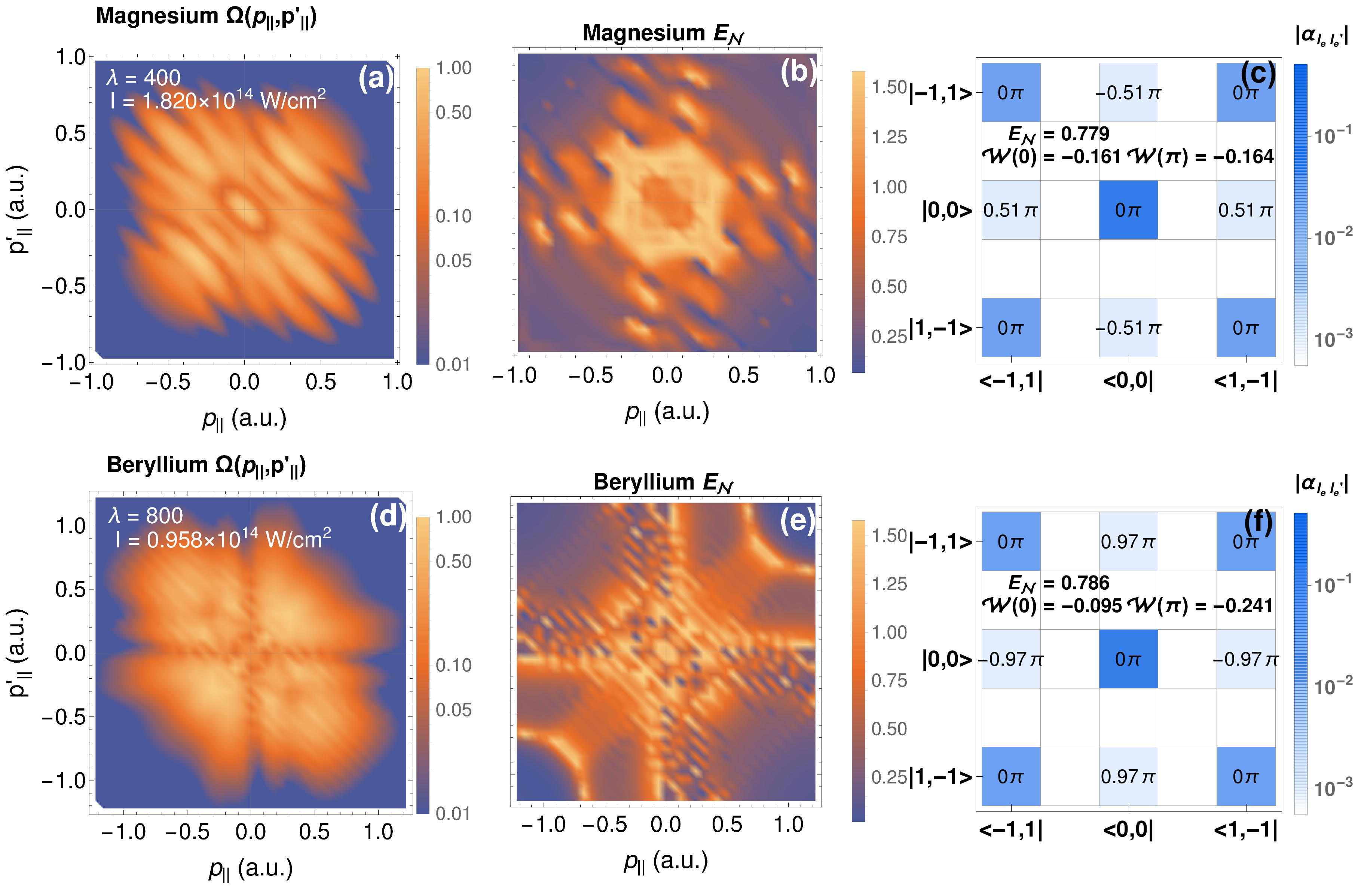}
		\caption{ First column, the correlated momentum distribution $\Omega(p_{||}, p'_{||})$, where we have integrated over the components perpendicular to the laser field polarization, given by \eqref{eq:correlated_dist}. The target is given in the title, while the field parameters are listed in the top left. In the middle row is the momentum-dependent logarithmic negativity (MDLN), where the perpendicular components are taken to be (nearly) zero, see \eqref{eq:MDLN} in \appref{sec:Methods:Entanglement:LogNeg} for more details. The right most column displays density matrices, $\rho_{OAM}$ \eqref{eq:Density_OAM}, for the indicated targets. The complex element of $\rho_{OAM}$, are represented by the phase $\arg(\alpha_{l_e l_e'})$ (recall that $\theta_{\alpha}=\arg(\alpha_{0 1})$) and the modulus $|\alpha_{l_el_e'}|$, see color bar. The bra and ket labels, panels (c) and (f), correspond to the states $\ket{l_e,l'_e}$, where the $l_e$ and $l'_e$ is the OAM of the two photoelectrons. 
		}
		\label{fig:DensityLogNeg}
	\end{figure*}
	
	In the panels (b) and (c), labeled $\mathcal{W}(0)$ and $\mathcal{W}(\pi)$, the expectation values of the corresponding entanglement witnesses are shown. Values above the zero line mean the entangled state may not be distinguished from separable states, while those below can. The witness $\mathcal{W}(\pi)$ outperforms $\mathcal{W}(0)$, which can be traced back to phases between the states $\ket{\pm1,\mp1}$ and $\ket{0,0}$ in the density matrix $\rho_{OAM}$ (denoted $\theta_{\alpha}$), which vary with the target and field parameters. The witness $\mathcal{W}(\pi)$ improves the detectability of entanglement for $\lambda=\SI{800}{\nm}$, while less difference is seen for $\lambda=\SI{400}{\nm}$.  For the example of beryllium in panel (e) $\theta_{\alpha}\approx\pi$ so the witness $\mathcal{W}(\pi)$ is much lower than $\mathcal{W}(0)$. However, for beryllium at $\lambda=400$~nm  and $I=3.8\times10^{14}$~W/cm$^2$ (the same ponderomotive energy) $\theta_{\alpha}\approx0.3\pi$ and the situation reverses $\mathcal{W}(0)<\mathcal{W}(\pi)$. Furthermore, for the example of magnesium in panel (d) at $\lambda=400$~nm $\theta_{\alpha}\approx0.5\pi$, which is why $\mathcal{W}(0)$ and $\mathcal{W}(\pi)$ give nearly identical results. While, for magnesium with $\lambda=800$~nm and $I=0.5\times10^{14}$ (the same ponderomotive energy) the phase is $\theta=0.9\pi$ and $\mathcal{W}(\pi)<<\mathcal{W}(0)$.
	Thus, the ability to measure entanglement depends on the phase between $\ket{0,0}$ and $\ket{\pm1,\mp1}$ in $\rho_{OAM}$, as well as a balanced distribution superposition of excited states with different $m_{\eta}$. However, computing this phase in advance allows the witness to be tuned, via the parameter $\theta$, to better suit the target and field parameters.
	
	In \figref{fig:DensityLogNeg}, we plot the correlated momentum distribution $\Omega(p_{||}, p'_{||})$ over $p_{||}$ and $p'_{||}$, where we have integrated over the components perpendicular to the laser field polarization. 
	\begin{equation}
		\Omega(p_{||}, p'_{||})\propto \iint \dd p_{\bot}\dd p'_{\bot}  p_{\bot} p'_{\bot} |M(\pv,\pvp)|^2 .
		\label{eq:correlated_dist}
	\end{equation}
	Interferences can be seen, which is a hallmark of the superposition of excited states \cite{hao_quantum_2014,maxwell_quantum_2015,maxwell_controlling_2016,quan_quantum_2017}. In the second column, we plot momentum-dependent logarithmic negativity (MDLN). Here, the logarithmic negativity is computed at specific momentum values, see caption and \appref{sec:Methods:Entanglement:LogNeg} for details, quantifying the entanglement between the photoelectrons given a specific final momentum. This can be somewhat related to the total logarithmic negativity, seen in \figref{fig:EntanglementSearch}, i.e., large areas with high MDLN and emission probability will lead to a large total logarithmic negativity.

	The MDLN varies considerably, forming `entanglement fringes', which follow the fringes in the momentum distribution. This trend is due to the fact that the momentum-dependent probability coherently mixes all values of $m_{\eta}$ and the resulting interference is related to the population between excitation channels of differing $m_{\eta}$. 
    For the MDLN of magnesium, panel (b), there are entanglement maxima either side of the $p_{||}$ and $p'_{||}$ axes, with diagonal fringes and a central ring of high logarithmic negativity.
	This behavior can be explained through the excitation channels  $4s_0$, $4f_0$ and $4f_{\pm1}$, which contribute the most, see \figref{fig:EntanglementSearch}. The combination of the three $f$ channels leads to the off axis maxima, as $4f_0$ has a node on the axis reducing the level of entanglement on axis but leading to a perfectly balanced superposition of channels with $m_{\eta}=-1$, $0$ and $1$ adjacent to the $p_{||}$ and $p'_{||}$ axes. However, the combination of $4f_0$ and $4f_{\pm1}$ also leads to the large diagonal fringes, similar structure are visible in both panel (a) and (b). The $4s$ channel has signal concentrated in the central region, around the $4f_0$ node in the center, this leads to a ring of high entanglement as a balanced OAM superposition is reached, but in the very center the $4s$ state begins to dominate, thus reducing the entanglement level.
	The MDLN of beryllium, panel (e), has the same entanglement maxima either side of the $p_{||}$ and $p'_{||}$ axes as magnesium. This arises from the states $4d_0$, $4d_{\pm1}$, $4f_0$ and $4f_{\pm1}$, which leads to the same effect due to on-axis nodes. In this case, however, the `entanglement fringes' are much narrower, which can also be seen in panel (d). Furthermore, the mixing of dominant channels with different $l$ leads to the large peaks on the diagonal (around $p_{||}, p'_{||} \approx\pm0.8$ a.u.), where the photoelectrons have large correlated momentum. 
	
	Through the interplay of excitation channels, we have identified three cases where the two photoelectrons approach the maximum logarithmic negativity. (i) Combination of three channels with equal $\ell$ and $m_{\eta}=\pm1,0$, which results in off-axis maxima, i.e., `fast' and `slow' entangled photoelectrons. (ii) The combination of an $s$ state with higher $\ell$ channel with $m_{\eta}=\pm1$, leads to central maxima or two `slow' entangled photoelectrons. (iii) The combination of two sets of differing $\ell$ channels each with $m_{\eta}=\pm1,0$, this leads to two `fast' pairs of entangled photoelectrons, with a \emph{correlated} direction.
	The different types of pairs of highly entangled photoelectrons could be useful as imaging probes accessing different momentum regions, or alternatively as a source of OAM entangled electrons, which can be optimized by tuning the interplay between different channels of excitation, for instance with tailored fields.
	
	In \figref{fig:DensityLogNeg}, we display $\rho_{OAM}$ for beryllium and magnesium targets. The phases of the complex entries of $\rho_{OAM}$ are written on each element, it is clearly only $\theta_{\alpha}$, the phase between $\ket{\pm1,\mp1}$ and $\ket{0,0}$, that plays a role. The closer the tuning parameter $\theta$ is to $\theta_{\alpha}$ the more effective the entanglement witnesses. The measure $E_{\mathcal{N}}$, on the other hand, is independent of such phases and is determined by the relative magnitude of the non-zero elements of $\rho_{OAM}$. We can see in the figure, it is the elements related to $\ket{\pm1,\mp1}$ and $\ket{0,0}$ that are most reduced. The coherence between these states is reduced after tracing over the momentum coordinates. The coherence between $\ket{\pm1,\mp1}$ and $\ket{\mp1,\pm1}$ is robust, however, as the channels that leads to these states such as $4d_{\pm1}$ are degenerate, so will have the same behavior over momentum. Thus, the final mixed state keeps a reasonably high logarithmic negativity.
	
	\section{Measurement Considerations}
	\label{sec:Results:Experimental}
	
	\begin{figure}
		\centering
		\includegraphics[width=0.49\textwidth]{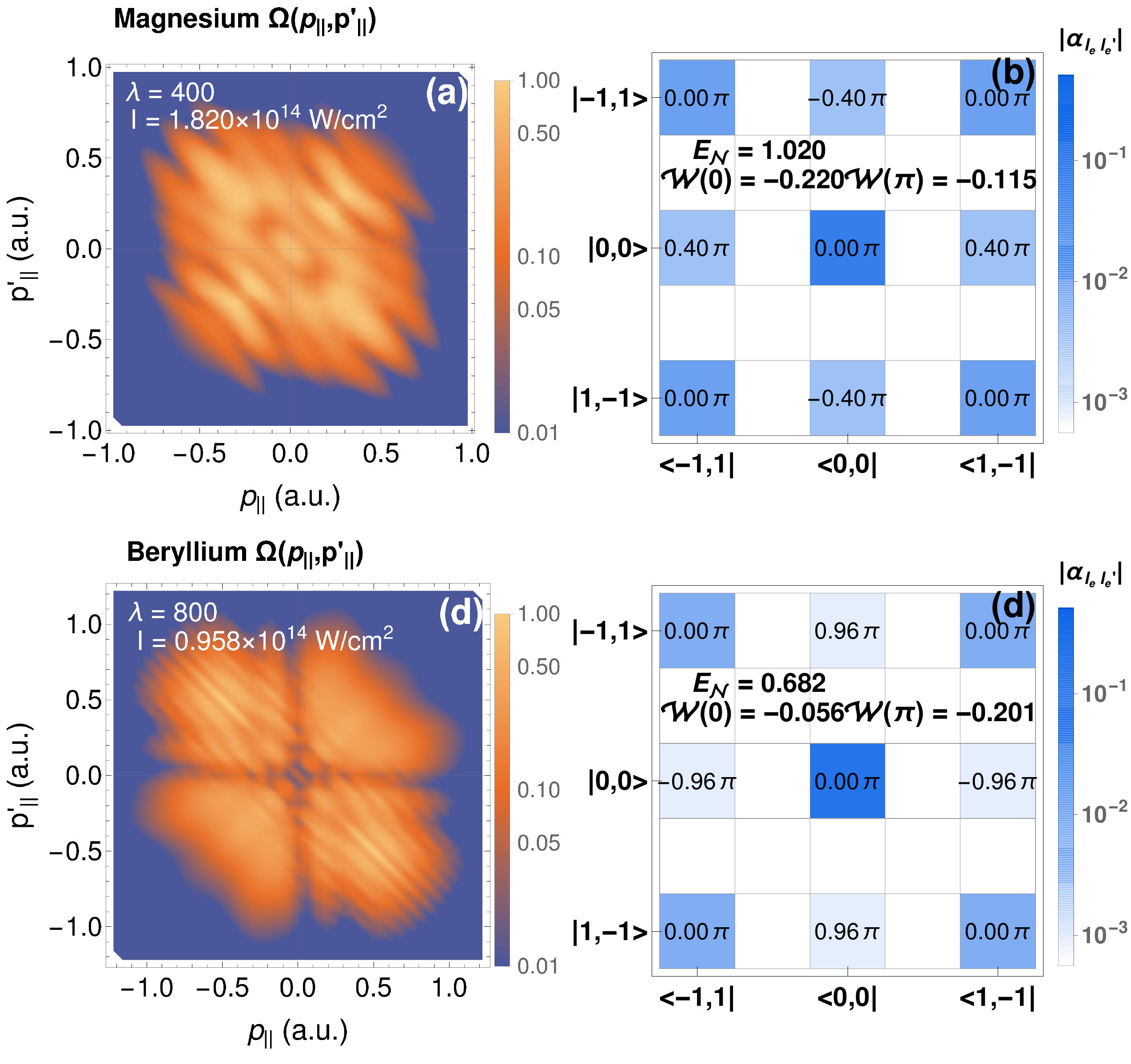}
		\caption{
		Same as \figref{fig:DensityLogNeg} but now the focal volume of the laser is accounted for. Focal averaged distributions, panels (a) and (c), and focal averaged density matrices, panels (b) and (d). The logarithmic negativity and entanglement witness values are recalculated and printed on the top right of panels (b) and (d).
		We use the focal averaging technique from \appref{sec:Methods:FocalAveraging}.  
		}
		\label{fig:FocalResults}
	\end{figure}

	In real strong-field experiments, there are additional incoherent averaging effects leading to mixed states. This will primarily be in the form of fluctuations and averaging over the carrier envelope phase and the laser intensity. For simplicity, we are assuming long enough pulses for the former to not be an issue. The latter takes place through intensity variation over focal volume (focal averaging) as well as intensity variation from shot-to-shot. We focus on focal averaging, the largest effect, following the procedure set out in Refs.~\cite{maxwell_controlling_2016,kopold_channelclosing_2002} and \appref{sec:Methods:FocalAveraging}. 
	
	In \figref{fig:FocalResults}, we show the results following focal averaging the momentum distribution as well as the density matrix. As expected, the interference fringes are reduced in the momentum distributions.	As an example, the central peak for magnesium is softened, in \figref{fig:DensityLogNeg}(a) the ratio between the peak and adjacent troughs is $\approx 6$ but after focal averaging, in \figref{fig:FocalResults}(a), this ratio is $\approx3$.
	But what effect does this have on the entanglement? Interestingly, the logarithmic negativity, labelled on the density matrix in panels (b) and (d), stays relatively high. In fact, for magnesium, it slightly increases.	 
	The reason for this is because, in the case of magnesium, the SFA predicts that at lower intensities the logarithmic intensity increases to $E_{\mathcal{N}}=1.06$ quite rapidly as lower energy excitation channels ($3p$ and $3d$) begin to play a role.
	We did not focus on these lower intensities due to their higher Keldysh parameter, outside our search region. In panel (b), the elements $\ket{0,0}$ and $\ket{\pm1,\mp1}$ increase in magnitude and there is a change in the phase associated with more contribution from these additional states. In contrast, in panel (d), we see a small reduction in the same element and the phase remains the same.
	In general, the elements associated with $\ket{\pm1,\mp1}$ and $\ket{\mp1,\pm1}$ remain coherent, so the entanglement is also robust to incoherent averaging over intensity.

	
	In its current form, \eqref{eq:Witness}, the expectation value of the witness can not be easily measured as it would require a combined measurement on both particles.
	However, it has been shown any witness of this type may be decomposed \cite{lewenstein_optimization_2000} into a series of \textit{local} measurements performed on each particle separately \cite{guhne_experimental_2003,bourennane_experimental_2004,bourennane_four_2004}. Practically, this will mean doing multiple experiments to compute expectation values and then combining these results with suitable weights determined by the decomposition. The sign of the combined results will determine if the electrons are entangled.
	For the specific values of $\theta=0$ and $\theta=\pi$, the witness decomposition is
	\begin{align}
    	&\mathcal{W}(0/\pi)=\notag\\&
    	\frac{1}{12}
    	\Bigg[
    	\frac{8}{3}\mathbb{1}^{\otimes 2}
    	\!-\!\lambda_3^{\otimes 2}
    	\!-\!2(
    	\lambda_4^{\otimes 2}
    	\!+\!\lambda_5^{\otimes 2}
    	)
    	\!+\!\lambda_8^{\otimes 2}
    	\!+\!\sqrt{3}(\lambda_3\!\otimes\!\lambda_8
    	\!+\!\lambda_8\!\otimes\!\lambda_3
    	)
    	\notag\\&
    	\hspace{1cm}\pm 2(
    	\lambda_1\!\otimes\!\lambda_6
    	\!+\!\lambda_6\!\otimes\!\lambda_1
    	\!+\!\lambda_2\!\otimes\!\lambda_7
    	\!+\!\lambda_7\!\otimes\!\lambda_2
    	)
    	)
    	\Bigg],
	\end{align}
	where $-$ is taken for $\theta=0$ and $+$ is taken for $\theta=\pi$.
	Here, $\mathbb{1}$ corresponds to the identity, i.e., performing no measurement, while $\lambda_i$ are matrices that label $8$ different single-particle measurements. 
	In the case of a two-level system (qubit/ SU(2)) the local measurement may correspond to the Pauli matrices plus the identity, which have a simple geometrical interpretation. 
	For our three-level system (qutrit/ SU(3)) we have used the eight Gell-Mann matrices. A geometrical interpretation for the Gell-Mann matrices can be found in Ref.~\cite{li_geometry_2013}. 
	As an example, $\lambda_3=L^{(1)}_{||}$ corresponds to a direct OAM measurement, while $\lambda_4$ corresponds to measuring in the basis ($\ket{-1}\pm\ket{1}$) and $\ket{0}$.
	For a full definition of the Gell-Mann matrices,
	see \appref{sec:Methods:Entanglement:Decomposition} and Ref.~\cite{gutierrez-esparza_experimental_2012}.
	
	We can use the decomposition \eqref{eq:WitnessDecomposition} to determine how it could reduce the total number of measurements required. We can do this by counting the number of combinations of $\lambda_x\otimes\lambda_y$, of which there are $11$. However, we can discount $\mathbb{1}^{\otimes 2}$ as this corresponds to doing nothing. Furthermore, $[\lambda_3, \lambda_8]=0$ (i.e. they share eigenstates) so all parts containing  only $\lambda_3$ and $\lambda_8$ can be determined with the same measurement. This leaves $7$ combinations, however measurements of the form $\lambda_x\otimes\lambda_y$ and $\lambda_y\otimes\lambda_x$ may be collected simutaneoulsy, as we do not distinguish the two electrons. This leaves a total of $5$ measurement settings, a large reduction to the 81 combinations of measurement for a general qutrit system. 
	See Ref.~\cite{inoue_measuring_2009} for an experimental example of full state tomography of all 81 measurement combinations of an OAM qutrit in photons, where an entanglement witness was not used.
	In contrast, see Ref.~\cite{gutierrez-esparza_experimental_2012}, for the use of an entanglement witness, where only 10 measurement setting were required, in an experiment to demonstrate entanglement for a photonic qutrit.
	%
	

	\section{Discussion}
	\label{sec:Discussion}
	In this work we use the correlated process of non-sequential double ionization (NSDI), as a backbone, to generate entanglement in the orbital angular momentum (OAM) of two photoelectrons. The photoelectron OAM in NSDI has some general and simple rules, which can easily be applied to show that entanglement only occurs in the recollision with subsequent ionization (RESI) mechanism of NSDI, prevalent at lower intensities, this is in agreement with Ref.~\cite{ho_nonsequential_2005} that classical correlation is a sufficient description for higher intensities. The OAM in NSDI has not been explored before, and previous entanglement studies have focused on the momentum coordinates parallel to the laser field \cite{liu_correlation_1999}. We find that OAM entanglement can be quite high approaching the maximally entangled state in specific momentum regions, and that this can be controlled by the interplay of channel of excitation. This could be investigated as a source of OAM-entangled electrons.
	Furthermore, the entanglement is robust as it survives incoherent averaging  over the focal volume. In addition, the entanglement demonstrated here is not generated by the symmetrization resulting from the indistinguishability of the photoelectrons \cite{schliemann_quantum_2001}.
	
	The use of the OAM has many benefits, firstly, the entanglement is simply understood as a consequence of angular momentum sharing during recollision, coupled with a superposition over OAM states due to contribution from excitation-channels with a differing quantum magnetic number [\figref{fig:NSDI_process}(c)]. Secondly, the quantization enables a clear 
	and simple analysis using the logarithmic negativity, which enables an extensive search over targets and parameters to maximize the entanglement, where it is clear ideal targets are those with two $s$-state valence electrons, given that this will promote a balenced superposition across OAM states [\figref{fig:EntanglementSearch}]. 
	Thirdly, the reduction in computational difficulty allows the density matrix and momentum-dependent logarithmic negativity to be computed [\figref{fig:DensityLogNeg}] and enables incoherent averaging [\figref{fig:FocalResults}], which all provide unique insight into the entanglement.
	Finally, we can construct an entanglement witness, which may be decomposed into local measurements, avoiding full state tomography or the measurements of incompatible continuous observables, like position and momentum, reducing the difficulty of experimental implementation for the detection of entanglement.
	
	A key questions is how to perform such an experiment? Firstly, we consider the measurement of OAM in attosecond experiments. The measurement of OAM for electron vortex beams, has received a great deal of attention and a variety of methods have been demonstrated \cite{saitoh_measuring_2013, guzzinati_measuring_2014,harvey_sterngerlachlike_2017,lloyd_electron_2017,grillo_measuring_2017,larocque_twisted_2018}. Most promising is the so-called OAM sorter \cite{grillo_measuring_2017, larocque_twisted_2018}, this has been proposed \cite{rotunno_orbital_2019,troiani_efficient_2020} and experimentally demonstrated \cite{tavabi_experimental_2021} as an OAM analyzing element to an electron microscope. Thus, such an element could be conceivably added to a typical reaction microscope (ReMi) \cite{moshammer_4p_1996,dorner_cold_2000,ullrich_recoilion_2003} employed for NSDI---a ReMi measures correlation of the ion and the two electrons, thus, it is already well-suited for studying entanglement \cite{schmidt-bocking_coltrims_2021}.
	To detect entanglement, the OAM sorter must be tuned to perform incompatible measurements on either electron independently. Such correlated measurements have been employ in detection of OAM-entangled light \cite{mair_entanglement_2001,inoue_measuring_2009,krenn_orbital_2017}.
	
	Beyond direct measurement, the entangled OAM of the photoelectrons in NSDI may be exploited in various ways, e.g., through interferometric schemes exploiting tailored fields, as a source of OAM-entangled electron pairs, or correlated measurements of OAM and momenta between the two electrons to disentangle interferences. 
	An interesting and unexplored consideration is the spin of the electrons and the spin-orbit coupling that may occur during recollision, as well as potential entanglement in the spin  or total angular momentum. OAM in attosecond processes provides a rich new research area, to help achieve the aim of imaging and controlling matter on ultrafast times scales and photoelectron OAM-entanglement plays an important role in achieving this goal. Beyond this, like Ref.~\cite{lewenstein_generation_2021}, it further demonstrates the fundamental non-classicality of such processes.

	
	
	
	
\section*{Acknowledgements}
	There were many interesting and useful discussion that predicated this study and thus, we would to thank, Prof.\ Carla Faria, Prof.\ Sougato Bose, Prof.\ Alessio Serafini, Prof.\ Jens Biegert, Prof.\ Misha Ivanov, Prof.\ Olga Smirnova and Dr.\ Emilio Pisanty.
	
	ASM acknowledges funding support from 
	the European Union’s Horizon 2020 research and innovation programme under the Marie Sk\l odowska-Curie grant agreement SSFI No.\ 887153. LBM acknowledges support from the Danish Council for Independent Research (Grant Nos.\ 9040-00001B and 1026-00040B).
	ML and ASM acknowledge support from ERC AdG NOQIA, State Research Agency AEI (“Severo Ochoa” Center of Excellence CEX2019-000910-S) Plan National FIDEUA PID2019-106901GB-I00 project funded by MCIN/ AEI /10.13039/501100011033, FPI, QUANTERA MAQS PCI2019-111828-2 project funded by MCIN/AEI /10.13039/501100011033, Proyectos de I+D+I “Retos Colaboración” RTC2019-007196-7 project funded by MCIN/AEI /10.13039/501100011033, Fundació Privada Cellex, Fundació Mir-Puig, Generalitat de Catalunya (AGAUR Grant No. 2017 SGR 1341, CERCA program, QuantumCAT \ U16-011424, co-funded by ERDF Operational Program of Catalonia 2014-2020), EU Horizon 2020 FET-OPEN OPTOLogic (Grant No 899794), and the National Science Centre, Poland (Symfonia Grant No. 2016/20/W/ST4/00314), Marie Sk\l odowska-Curie grant STREDCH No 101029393, “La Caixa” Junior Leaders fellowships (ID100010434),  and EU Horizon 2020 under Marie Sk\l odowska-Curie grant agreement No 847648 (LCF/BQ/PI19/11690013, LCF/BQ/PI20/11760031,  LCF/BQ/PR20/11770012).).

\appendix		
	

	\section{Strong-Field Approximation}
	\label{sec:Methods:Models:SFA}
	
	The wave function for NSDI, which describes both electron impact ionization (EI) and recollision with subsequent ionization (RESI) can be written using an SFA flavor ansatz \cite{amini_symphony_2019} in the following way
    \begin{align}
    	\ket{\psi(t)}&=
    			a(t)\ket{0}
    			\hspace{-1mm}+\hspace{-1mm}\int \dd^3 \pv\, b(\pv, t)\ket{\pv,0}\Bigg.
    			\hspace{-1mm}+\hspace{-1mm}\sum_{\eta}\!\! \int \dd^3 \pv\, c(\pv,\eta, t)\ket{\pv,\eta}
    			\notag\\&\hspace{14mm}
    			+\iint\dd^3 \pv \dd^3 \pvp \, d(\pv,\pvp,t)\ket{\pv,\pvp},
    	\label{eq:RESI_Maciej}
    \end{align}
	where $\ket{0}$ is the two electron ground state, $\ket{\pv,0}$ corresponds to one electron in the continuum and the other in its ground state, $\ket{\pv,\eta}$ is the same as the latter with the second electron excited and $\ket{\pv,\pvp}$ is the two electron continuum state.
	
    The final symmetrized transition amplitude is related to \eqref{eq:RESI_Maciej} via
	\begin{align}
	M(\pv,\pvp)&=\lim_{t\rightarrow\infty}d(\pv,\pvp,t)\\
	&=\frac{1}{\sqrt{2}}\left(
	M_{\mathrm{unsym}}(\pv,\pvp)
	+M_{\mathrm{unsym}}(\pvp,\pv)
	\right) ,\notag
	\end{align}
	where $M_{\mathrm{unsym}}(\pv,\pvp)$ is the unsymmetrized transition amplitude. This is valid for an initial singlet state, as considered here.	
	In the SFA, in atomic units, for the RESI mechanism and using the assumptions listed in Refs.~\cite{maxwell_quantum_2015,maxwell_controlling_2016}, the unsymmetrized transition amplitude can be written as
	\begin{align}
	&M_{\mathrm{unsym}}(\pv,\pvp)=\notag\\
	&\sum_{\eta}\int \dd^3 t\int \dd^3 \kv 
	V_{\pv\pvp,\pv \eta}V_{\pv \eta,\kv 0}V_{\kv 0, 0}
	\exp [iS(\pb,\pb',\kb,t,t',t'')],  \label{eq:Mp}
	\end{align}
	where
	\begin{align}
	\int \dd^3 t&\equiv\int_{-\infty}^{\infty}\dd t\int_{-\infty}^{t}\dd t'\int^{t'}_{\infty} \dd t''
	\intertext{and}
	&S(\pb,\pb',\kb,t,t',t'')=
	\notag \\	& 
	I_{p}^{\mathrm{10}}t''+I_p^{\mathrm{20}}t'
	+I_p^{\mathrm{\eta}}(t-t')-\int_{t''}^{t'}
	\frac{[\kb+\Ab(\tau )]^{2}}{2}\dd\tau
	 \notag \\	& 
	 -\int_{t'}^{\infty }\frac{[\pb+\Ab(\tau )]^{2}}{2}\dd\tau
	-\int_{t}^{\infty }\frac{[\pb'+\Ab(\tau )]^{2}}{2}\dd\tau  \label{eq:singlecS}
	\end{align}
	denotes the semiclassical action and $I_p^{\mathrm{10}}$, $I_p^{\mathrm{20}}$ and
	$I_p^{\mathrm{\eta}}$ are the one-electron ionization potentials corresponding to removing a bound electron from $\ket{0}$, $\ket{\pv,0}$ and $\ket{\pv,\eta}$, respectively. 
	Note that arrow notation is reserved for three-dimensional vectors $\pv=(p_{||},p_{\perp},\phi)$, while bold face notations relates to two-dimensional vectors $\pb=(p_{||},p_{\perp})$, where $p_{||}=p_{z}$, $p_{\perp}=\sqrt{p_{x}^2+p_y^2}$ and $\phi$ is the azimuthal angle.
	The prefactors are given by
	\begin{align}
	V_{\kv 0, 0} &=\braket{{\kv}(t''),0| V| 0}  \notag \\
	&=\frac{1}{(2\pi )^{3/2}}\int \dd^{3}\rv\; V({\rv})e^{-i{\kv}(t'')\cdot \rv}\psi _{10}(\rv),  \label{eq:Vkg}\\
	V_{\pv \eta,\kv 0} &=\braket{ {\pv}(t') ,\eta| V_{12}| {\kv}(t'),0}  \notag \\
	&=\frac{1}{(2\pi )^{3}}\iint \dd^{3}\rvp\dd^{3}\rv\exp [-i(\pv
	-\kv)\cdot {\rv}]\notag\\
	& \hspace{1.75cm}\times
	V_{12}({\rv},{\rvp})[\psi _{\eta}({\rvp})]^{*}\psi_{20}({\rvp})\quad  \label{eq:Vp1e,kg}
	\intertext{and}
	V_{\pv\pvp,\pv\eta} &=\braket{ {\pv}( t),{\pv}\;'( t)| V_{\mathrm{ion}}| {\pv}( t), \eta}  \notag\\
	&=\frac{1}{(2\pi )^{3/2}}\int \dd^{3}\rvp 
	V_{\mathrm{ion}}({\rvp})e^{-i{\pv}\;'(t)\cdot {\rvp}}\psi_{\eta}({\rvp}),  \label{eq:Vp2e}
	\end{align}
	where ${\pv}(t)$, ${\pvp}(t)$ and ${\kv}(t)$ are defined according to ${\kv}(t)=\kv+\Av(t)$ or ${\kv}(t)=\kv$ in the length or velocity gauge, respectively. In this work, for simplicity, we employ the velocity gauge.
	This formalism describes the RESI process, in which an electron is ionized by the laser field from the ground state $\ket{0}$ at time $t''$ into $\ket{\kv, 0}$ with intermediate momentum $\kv$, it recollides at $t'$ and excites a second electron into the state $\ket{\pv, \eta}$ with a final momentum $\pb$ for the initial electron. The second electron is ionized via the laser field at time $t$ into the state $\ket{\pv, \pvp}$ with a final momentum $\pvp$.
	The prefactors give the information about all the bound states \cite{shaaran_laserinduced_2010,shaaran_laserinduced_2010a} and interactions, for which we employ $V$ the singly charged binding potential for the first electron, $V_{\mathrm{ion}}$ the doubly charged binding potential for the second electron and $V_{12}$ the electron--electron interaction.
	In this approximation, electron--electron correlation is described by the prefactor $V_{\pv \eta,\kv 0}$.
	The transition amplitude (\ref{eq:Mp}) is computed using the steepest descent method, as described in Ref.~\cite{figueirademorissonfaria_highorder_2002}.
	In which, we look for values of the variables $t$, $t'$ $t''$ and $\kv$ such that the action is stationary. This leads to the following saddle-point equations
	\begin{align}
	\left[ \kv+\Av(t'')\right] ^{2}&=-2I_p^{\mathrm{10}},
	\label{eq:saddle1}\\
	\kv&=-\frac{1}{t'-t''}\int_{t''}^{t'}\dd\tau \Av(\tau ),
	\label{eq:saddle2}\\
	\left[ \pv+\Av(t^{\prime })\right]^2
	&=\left[\kv+\Av(t')\right]^{2}
	-2(I_p^{\mathrm{20}}-I_p^{\mathrm{\eta}}),
	\label{eq:saddle3}\\
	\intertext{and}
	\left[ \pvp+\Av(t)\right]^{2}&=\mathbf{-}2I_p^{\mathrm{\eta}}.
	\label{eq:saddle4}
	\end{align}
	Eqs.~(\ref{eq:saddle1}) and (\ref{eq:saddle4}) give the energy conservation of the first and second electron at the instant of tunnel ionization, \eqref{eq:saddle2} enforced the first electron will return and \eqref{eq:saddle3} describes energy sharing between the electrons.
	
	\section{Conservation Laws}
	\label{sec:Methods:Models:Conservation}
	The conservation laws exploited in the main text are general and can be arrived at with few assumptions.
	The Hamiltonian of the two-electron system may be written as
	\begin{equation}
    	    \hat{H}(t)=\frac{\hat{p}^2\!+\!{\hat{p}'}{^2}}{2}
    	    \!+\!(r_{||}\!+\!r'_{||}) E_{||}(t)
	    \!+\!V_{ion}(\rv)\!+\!V_{ion}(\rvp)
	    \!+\!\hat{V}_{12}.
	\end{equation}
	All terms are independent of the coordinates $\phi$ and $\phi'$ except for $\hat{V}_{12}$, which depends on the relative distance between the electrons $|\rv-\rvp|=\sqrt{(r_{||}-r_{||}')^2+r_{\perp}^2{r'}_{\perp}^2\cos(\phi-\phi')}$, thus this is still invariant to a rotation to both particles. Hence, $[\hat{L}_{||},\hat{H}]=0$, where $\hat{L}_{||}=-i\partial_{\phi}-i\partial_{\phi'}$, and total OAM is conserved, $m+m'=l+l'$. If, during ionization by the laser field, the single-active-electron approximation is employed (as is widespread) orbital angular momentum will be conserved during ionization, given that $[\hat{L}^{(1)}_{||},\hat{H}^{(1)}]=[\hat{L}^{(2)}_{||},\hat{H}^{(2)}]=0$, where $\hat{L}^{(1)}$, $\hat{L}^{(2)}$, $\hat{H}^{(1)}$ and $\hat{H}^{(2)}$ are the OAM operators and Hamiltonians for each individual electron, including only one-particle terms. Thus, this lead to $l'=m_{\eta}$. 
	
	The same conservation laws plus additional constraints are encoded in the SFA via the prefactors, which may be written in terms of their dependence on the azimuthal angles
	\begin{align}
	    V_{\kv 0,0}&=e^{i m\phi_{\kb}}\tilde{V}_{\kb 0,0}\\
	    V_{\pv \eta, \kv 0}&=
	    e^{i (m'+m_{\eta})\phi_{\pb}}\tilde{V}_{\pb \eta, \pb'' 0}\\
	    V_{\pv \pvp,\pv \eta}&=
	    e^{i m_{\eta}\phi_{\pb'}}\tilde{V}_{\pb \pb',\pv \eta},
	\end{align}
	where the tilde indicates quantities independent of the azimuthal angle of all coordinates.
	Now using \eqref{eq:2dFourier} and the above equations we may write the SFA OAM transition amplitude
	\begin{align}
	    &M_{l_e,l_e'}^{RESI}(\pb,\pb')=i^{-(l_e+l_e')}
	    \delta_{m,0}
	    \delta_{m'-m_{\eta},l_e}
	    \delta_{m_{\eta},l_e'}
	    \tilde{M}(\pb, \pb')
	    \intertext{with}
	    &\tilde{M}(\pb, \pb')\!=\!\int\! \dd^3 t\!\int\! \dd^2 \kb 
	\tilde{V}_{\pb\pb',\pb \eta}
	\tilde{V}_{\pb \eta,\kb 0}
	\tilde{V}_{\kb 0, 0}
	e^{ iS(\pb,\pb',\kb,t,t',t'')}.
	\end{align}
	Thus, we recover the above stated conservation equations along with the condition $m=0$. The second electron in its ground states has quantum magnetic number $m'$, and from the behavior of the spherical harmonic $Y^{\ell}_{-m'}(\Omega)=(-1)^{m'}Y^{\ell}_{m'}(\Omega)$ we can deduce that, for odd values of $m'$ we will get opposite signs in the final transition amplitude and thus odd pairs will cancel, as the initial states will be degenerate. Thus, for the $s$ and $p$ initial states employed here, we can assume $m'=0$.

	
	\section{Density Matrix}
	\label{sec:Methods:Entanglement:Density}
	
	The full density matrix $\rho=\ket{{\psi}}\bra{{\psi}}$, where $\ket{\psi}$ is given by \eqref{eq:OAM_Qutrit}, is
	\begin{widetext}
	\begin{align}
	\rho&=\sum_{l_e,l_e'}\!\iint \dd^2 \pb  \dd^2 \pb'  \dd^2 \pb''  \dd^2 \pb'''
	M_{l_e, -l_e}(\pb,\pb')  
	M_{l_e', -l_e'}^*(\pb'',\pb''')
	\ket{\pb,l_e,\pb',-l_e}\bra{\pb'',l_e',\pb''',-l_e'}.
	\end{align}
	\end{widetext}
	We do not compute this explicitly due to the continuous momentum coordinates, instead most commonly, we will trace out the momentum coordinates 
	\begin{equation}
	    \rho_{OAM}=\iint \dd^2 \kb\dd^2\kb' \bra{\kb,\kb'}\rho\ket{\kb,\kb'},
	    \label{eq:Density_Trace}
	\end{equation}
	where we assume $\braket{\kb|\pb,l_e}=\delta(\kb-\pb)\ket{l_e}$. Here $\ket{l_e}$ is an `OAM state' with the property $\braket{r|l_e}\propto e^{i l_e \phi}$. Applying this rule results in \eqref{eq:Density_OAM}.
	In order to compute the momentum dependent logarithmic negativity (MDLN) we need to compute the density matrix conditioned on some specific final momentum
	\begin{align}
	    &\rho(\pb,\pb')=\notag\\
	    &\sum_{l_e,l_e'}\!
	M_{l_e, -l_e}(\pb,\pb')
    M_{l_e', -l_e'}^*(\pb,\pb')
	\ket{\pb,l_e,\pb',-l_e}\bra{\pb,l_e',\pb',-l_e'}.
	\end{align}
	
	\section{Positive Partial Transpose}
	\label{sec:Methods:Entanglement:PPT}
	The positive partial transpose (PPT) or Peres-Horodecki criterion \cite{peres_separability_1996,horodecki_separability_1997} is a \emph{necessary} condition on density matrices to determine if a system is separable. It is valid for both pure and mixed states. The approach is to take the \textit{partial transpose}---i.e. transpose one subsystem---and compute the eigenvalues, if any are negative the state is non-separable and thus, entangled. For $2\otimes2$ and $2\otimes3$ the condition is also \emph{sufficient}, so no negative eigenvalues imply separability, for higher dimensional systems (such as our $3\otimes2$ system), this is not the case, however, it is still a very powerful method and witnesses can be constructed to detect any `PPT entangled' state \cite{lewenstein_characterization_2001}, which we can exploit to our advantage.
	
	For our NSDI qutrit mixed state
	\begin{equation}
    	\rho_{OAM}=\sum_{l_e,l_e'=-1}^{1}
    	\alpha_{l_e l_e'}\ket{l_e,-l_e}\bra{l_e',-l_e'},
    \end{equation}
    where $\alpha$ is defined as in \eqref{eq:Density_OAM}. In taking the partial transpose we swap the indices for one of the subsystems (in this case the `\textit{second}' electron),
    \begin{equation}
    	\rho_{OAM}=\sum_{l_e,l_e'=-1}^{1}
    	\alpha_{l_e l_e'}\ket{l_e,-l_e'}\bra{l_e',-l_e}.
    \end{equation}
    The eigenvalues can be computed analytically as
    \begin{align}
        \alpha_{l_e l_e}=\int\dd^2 \pb \int \dd^2 \pb'|M_{l_e-l_e}(\pb,\pb')|^2 && \text{for}&& l_e\in[-1,1],
    \end{align}
	which are positive and $\pm|\alpha_{l_e l_e'}|$ for $l_e\ne l_e'$, which provides three negative eigenvalues as long as $|\alpha_{l_e l_e'}|\ne0$ for $l_e\ne l_e'$. Thus, the electrons from the RESI mechanism of NSDI are always PPT entangled as long as there is non-zero population across different excited state with differing $m_{\eta}$. This also means there will always be an entanglement witness, which may be used to experimentally verify this entanglement.
	
	\section{Logarithmic Negativity}
	\label{sec:Methods:Entanglement:LogNeg}
	The logarithmic negativity is well-suited for quantifying entanglement in PPT entangled states, as it is equal to the `entanglement cost' to create this entanglement via PPT operations. In the main article we construct the logarithmic negativity from the reduced density matrix, traced over momentum coordinates, however, we may instead define a momentum dependent logarithmic negativity (MDLN)
	\begin{equation}
		E_{\mathcal{N}}(p_{||},p_{||}')=\log_2\left[\left|\left| \rho^{T_A}(p_{||},\delta p,p_{||}',\delta p)
		\right|\right|_1\right]\notag\label{eq:MDLN},
	\end{equation}
	where the perpendicular momentum coordinates are set to nearly to zero,  $\delta p=0.05$~a.u., in order allow for 2D visualization. The value is chosen to be where the distribution will have high probability but no nodes due to the geometry of the excited states.
	
	The logarithmic negativity can be related to the sum of the negative eigenvalues ($\lambda_i$) of the density matrix
	\begin{align}
		E_{\mathcal{N}}&=\log_2\left[|| \rho^{T_A}_{OAM} ||_1\right]\notag\\
		&=\log_2\left(1+
			2\left|\sum_{\lambda_i<0}\lambda_i\right|
		\right)
		\intertext{which can be written as minus the sum of the negativive eigenvalues}
		&=\log_2\left(
		1+2\left(|\alpha_{-1 0}|+|\alpha_{-1 1}|+|\alpha_{-1 0}|+|\alpha_{0 1}|\right)
		\right)
	\end{align}
	Using the Cauchy–Schwartz inequality we can show $\alpha_{l_e l_e'}\le \sqrt{\alpha_{l_e l_e}\alpha_{l_e' l_e'}}$, leading to
	\begin{align}
		&|\alpha_{-1 0}|+|\alpha_{-1 1}|+|\alpha_{-1 0}|+|\alpha_{0 1}|\notag\\
		&\le
		\sqrt{\alpha_{-1-1}\alpha_{00}}+\sqrt{\alpha_{-1-1}\alpha_{11}}+\sqrt{\alpha_{00}\alpha_{11}}\notag\\
		&\le\sqrt{\alpha_{-1-1}+\alpha_{11}+\alpha_{00}}\sqrt{\alpha_{00}+\alpha_{-1-1}+\alpha_{11}}\notag\\
		&= 1,
	\end{align}
	where other forms of the Cauchy–Schwartz inequality were used in the additional inequalities. With this we can place a bound upon the logarithmic negativity
	\begin{equation}
		E_{\mathcal{N}}\le \log_2(1+2)\approx 1.58\;.
	\end{equation}
	
	\section{Entanglement Witnesses}
	\label{sec:Methods:Entanglement:Witness}
	Here, we will show that the witness used in this work (see \eqref{eq:Witness}) is a valid entanglement witness. We do this by showing that (i) it is positive for all separable states and (ii) the trace is negative for at least one entangled state. For a general separable pure state
	\begin{equation}
	    \ket{\psi}=\sum_i a_i \ket{i} \otimes \sum_j b_i \ket{j}
	\end{equation}
	the density matrix may be written as
	\begin{equation}
		\rho_s=\sum_{i,j,m,n}a_i a^*_m b_j b^*_n \ket{i,j}\bra{m,n}
	\end{equation}
	and we can compute the trace with the witness of \eqref{eq:Witness}
	\begin{align}
		\mathrm{Tr}[\rho_s \mathcal{W}(\theta)]&=\frac{1}{d}
		\sum_{i,j}|a_i|^2|b_j|^2-\bra{\nu(\theta)}\rho_s\ket{\nu(\theta)}\\
		&=\frac{1}{d}-\left|\sum_{ij}a_i b_j\braket{i,j|\nu(\theta)}\right|^2\\
		&=\frac{1}{d}-\frac{1}{d}\left|\sum_{l_e} a_{l_e} e^{i\theta l} b_{-l_e}\right|^2\\
		&\ge \frac{1}{d}-\frac{1}{d}\sum_{l_e} |a_l e^{i\theta l_e}|^2
		\sum_{l_e} |b_{l_e}|^2
		=0.
	\end{align}
	Thus, we have demonstrated (i) and the entangled state we use for (ii) is $\ket{\nu(\theta)}$ 
	\begin{align}
	\mathrm{Tr}[\ket{\nu(\theta)}\bra{\nu(\theta)} \mathcal{W}(\theta)]&=\frac{1}{d}-|\braket{\nu(\theta)|\nu(\theta)}|^2\\
	&=-\frac{d-1}{d}<0.
	\end{align}
	Hence, given $d=3$, we have shown that we are using a valid witness.
	
	\section{Witness Decomposition}
	\label{sec:Methods:Entanglement:Decomposition}
	
	The decomposition of entanglement witness into a series of local measurement is described in Refs.~\cite{lewenstein_optimization_2000,guhne_experimental_2003,bourennane_experimental_2004,bourennane_four_2004}. As described in the main text, for a qutrit, a convenient decomposition is in terms of the Gell-Mann matrices.
	These are defined by the following construction \cite{gutierrez-esparza_experimental_2012},
	\begin{align}
		&\chi^{\pm}_{l_e l_e'}=\ket{x^{\pm}_{l_e l_e'}}\bra{x^{\pm}_{l_e l_e'}}
		&&\text{and}
		&&\ket{x^{\pm}_{l_e l_e'}}=\frac{1}{\sqrt{2}}(\ket{l_e}\pm\ket{l_e'}),
		\notag\\
		&\Upsilon^{\pm}_{l_e l_e'}=\ket{y^{\pm}_{l_e l_e'}}\bra{y^{\pm}_{l_e l_e'}}
		&&\text{and}
		&&\ket{y^{\pm}_{l_e l_e'}}=\frac{1}{\sqrt{2}}(\ket{l_e}\pm i \ket{l_e'}),
	\end{align}
	the matrices are then given by
	\begin{align}
		&\lambda_1=\chi^{+}_{01}-\chi^{-}_{01}		
		&&\lambda_2=\Upsilon^{+}_{01}-\Upsilon^{-}_{01}		
		\notag\\
		&\lambda_4=\chi^{+}_{0-1}-\chi^{-}_{0-1}		
		&&\lambda_5=\Upsilon^{+}_{0-1}-\Upsilon^{-}_{0-1}
		\notag\\
		&\lambda_6=\chi^{+}_{1-1}-\chi^{-}_{1-1}
		&&\lambda_7=\Upsilon^{+}_{1-1}-\Upsilon^{-}_{1-1}		
		\notag\\
		&\lambda_3=\ket{0}\bra{0}-\ket{1}\bra{1}
		\notag\\
		&\mathrlap{\lambda_8=\frac{1}{\sqrt{3}}(\ket{0}\bra{0}+\ket{1}\bra{1}-2\ket{-1}\bra{-1}).}
	\end{align}
	These matrices can be related to the well-known Pauli matrices.
	For example, the measurement $\lambda_3=\hat{L}_{||}$, corresponds directly to an OAM measurement, similar to the $\sigma_z$ Pauli matrix, while $\lambda_{1}$,  $\lambda_{4}$,  $\lambda_{6}$ can be related to $\sigma_x$ as they are all pairwise superposition of two OAM states with the phase $\pm$. The measurements $\lambda_{2}$,  $\lambda_{5}$,  $\lambda_{7}$ relate to $\sigma_y$ as they are pairwise superposition with the phase $\pm i$. Measurements $\lambda_1$--$\lambda_7$ have eigenvalues of $-1$, $0$, $1$, as with OAM, while $\lambda_8$ has eigenvalues $-2/\sqrt{3}$ and $1/\sqrt{3}$.
	
	The full entanglement witness decomposition given in terms of the Gell-Mann matrices is
	\begin{align}
		&\mathcal{W}(\theta)=
		\frac{1}{12}
		\Bigg[
			\frac{8}{3}\mathbb{1}^{\otimes 2}
			-\lambda_3^{\otimes 2}
			-\kappa^2(\kappa^{-4}\!+\!1)(
				\lambda_4^{\otimes 2}
				+\lambda_5^{\otimes 2}
			)
			+\lambda_8^{\otimes 2}
			\notag\\&
			+\sqrt{3}(\lambda_3\!\otimes\!\lambda_8
				\!+\!\lambda_8\!\otimes\!\lambda_3
			)
			+i\kappa^2 (\kappa^{-4}\!\!-\!1)(
				\lambda_4\!\otimes\!\lambda_5
				\!-\!\lambda_5\!\otimes\!\lambda_4
			)
			\notag\\&
			-\kappa (\kappa^{-2}\!+\!1)(
				\lambda_1\!\otimes\!\lambda_6
				\!+\!\lambda_6\!\otimes\!\lambda_1
				\!+\!\lambda_2\!\otimes\!\lambda_7
				\!+\!\lambda_7\!\otimes\!\lambda_2
			)
			\notag\\&
			-i\kappa (\kappa^{-2}\!-\!1)(
				\lambda_1\!\otimes\!\lambda_7
				\!-\!\lambda_7\!\otimes\!\lambda_1
				\!-\!\lambda_2\!\otimes\!\lambda_6
				\!+\!\lambda_6\!\otimes\!\lambda_2
			)
		\Bigg],
		\label{eq:WitnessDecomposition}
	\end{align}
	where $\kappa=e^{i\theta}$.
	This decomposition is specific to the Gell-Mann matrices, however, a general procedure may be outlined for any complete set of measurement bases. Thus, if it is simpler to measure in another way, use another entanglement witness or perform this for a higher-dimensional system, the following recipe can be used to achieve the decomposition. For a complete set of one-particle measurements defined by the operators $\lambda_i$ for $i\in[1,d^2]$ and a known entanglement witness $\mathcal{W}$, the decomposition in terms of the local one-particle observables may be written as
	\begin{equation}
	    \mathcal{W}=\sum_{ij} c_{ij} \lambda_i \otimes \lambda_j,
	    \label{eq:GeneralDecomposition}
	\end{equation}
	where the coefficients $c_{ij}$ can be used to combined with experimental local expectation values to determine $\mathcal{W}\!$. This sum can be inverted to find $c_{ij}$ via vectorization, i.e. flattening one dimension. We may define 
	\begin{align}
    	\mathcal{W}_v&=\mathrm{vec}(\mathcal{W})\notag\\
    	&=(\mathcal{W}_{0,0},\mathcal{W}_{0,1},...,\mathcal{W}_{0,d^2-1},\mathcal{W}_{1,0},...,\mathcal{W}_{d^2-1,d^2-1})^T
	\end{align}
	and 
	$c_v=(c_{0,0},c_{0,1},...,c_{0,d^2-1},c_{1,0},...,c_{d^2-1,d^2-1})^T$.
	This reduces the $d^2\times d^2$ matrix operators, $\mathcal{W}$ and $c$, to $d^4$ dimensional vectors. 
	Now we defined a  $d^4\times d^4$ matrix 
	\begin{align}
	    M_v=\left(
	    \mathrm{vec}(\lambda_0 \otimes \lambda_0),..
	    ,\mathrm{vec}(\lambda_{d-1} \otimes \lambda_{d-1})
	    \right),
	\end{align}
	where each row is the vectorized matrix $\lambda_i \otimes \lambda_j$ for specific values of $i$ and $j$. Now using these definitions we may rewrite \eqref{eq:GeneralDecomposition} as a linear matrix equation $\mathcal{W}_v=M c_v$, thus we may obtain the coefficients by inverting, such that $c_v=M^{-1} \mathcal{W}_v$. For our finite dimensional system with $d=3$, this can be done very quickly and straightforwardly.
	
	\section{Focal Averaging}
	\label{sec:Methods:FocalAveraging}
	The focal averaging used for the momentum-dependent probability distributions has been discussed extensively in Ref.~\cite{kopold_channelclosing_2002} and our previous work \cite{maxwell_controlling_2016}. The extension to density matrices is new, but takes an almost identical form. The basic equations for focal averaging are given by integrating the ionization rate for a specific intensity over the focal volume and duration of the laser field to get something proportional to the number of electrons measured at a specific momentum in experiment
	\begin{equation}
	    N(p_{\parallel}, p'_{\parallel})\propto\int\dd t\int \dd^3\rb\Omega(p_{\parallel},p'_{\parallel},I(r_{||}, r_{\perp}, t)),
	    \label{eq:FocalGen}
	\end{equation}
	where $\Omega(p_{\parallel},p'_{\parallel},I)$ is the probability $|M(p_{\parallel},p'_{\parallel})|^2$ given at a specific intensity $I$ and  $I(r_{||}, r_{\perp}, t)$ gives an approximation to the laser beam intensity profile
	\begin{equation}
	    I(r_{||}, r_{\perp}, t) = I_0\frac{d_0}{d(r_{||})}  
	    \exp\left(-\frac{2 r_{\perp}^2}{d(r_{||})^2} \right)
	    \exp\left(\frac{(t-r_{||}/c)^2}{\tau^2}\right)
	\end{equation}
	with $d(z)=d_0\left[1+(z/z_0)^2\right]^{\frac{1}{2}}$, where $d_0=\sqrt{\lambda z_0/\pi}$ is the beam waist, $I_0$ is the peak intensity, $z_0$ is the Rayleigh length, and $c$ is the speed of light. Then \eqref{eq:FocalGen} may be simplified and parameterized in terms of an integral over the laser intensity
	\begin{align}
	   N(p_{\parallel},p'_{\parallel})&\propto 
	   \int_0^{I_0}\dd I f(I) \Omega(p_{\parallel},p'_{\parallel},I),
	   \intertext{where}
	   f(I)&=\frac{1}{I}\int^{\eta}_0 \dd\eta (1+\eta^2)\ln\left( \frac{I_0}{I(1+\eta^2)} \right)^{1/2}
	\end{align}
	
	In order to take into account the focal volume for density matrices, a similar process must be performed. We can write the density matrix corresponding to a specific laser intensity as $\rho(I)$, now the mixed state that accounts for all intensities over the focal volume is the weighted sum/ integral over all intensities in the volume, where the weights are given by $f(I)$ such that
	\begin{equation}
	    \rho_{\mathrm{focal}}\propto\int_0^{I_0}\dd I f(I) \rho(I).
	\end{equation}
	
    \bibliography{NSDI_OAM}

\end{document}

%% file: SF_Def.tex
\newcommand{\figref}[1]{\figurename~\ref{#1}}
\newcommand{\secref}[1]{Sec.~\ref{#1}}
\newcommand{\appref}[1]{Appendix \ref{#1}}
\renewcommand{\eqref}[1]{Eq.~(\ref{#1})}

\newcommand{\ip}{{I}_{p}}
\newcommand{\up}{{U}_{p}}
\newcommand{\pb}{\mathbf{p}}
\newcommand{\pv}{\vec{p}}
\newcommand{\pvp}{\vec{p}^{\,\prime}}

\newcommand{\kb}{\mathbf{k}}
\newcommand{\kv}{\vec{k}}
\newcommand{\dd}{d}
\newcommand{\rb}{\mathbf{r}}
\newcommand{\rv}{\vec{r}}
\newcommand{\rvp}{\vec{r}^{\,\prime}}

\newcommand{\Ab}{\mathbf{A}}
\newcommand{\Av}{\vec{A}}

\DeclareSIUnit{\au}{{a.u.}}
\DeclareSIUnit{\nm}{{nm}}

\newcommand{\orcid}[1]{%
  ~\href{https://orcid.org/#1}{\includegraphics[width=8pt]{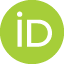}}%
  }